\documentclass[superscriptaddress,amsmath,amssymb,aps,twocolumn,prc]{revtex4-1}
\usepackage{graphicx}
\usepackage{color}

\def\nonubb{$\beta\beta(0\nu)$}

\def\enrge{${}^{\mathrm{enr}}$Ge}

\def\gae{$g_{ae}$}
\def\gaegan{$g_{ae}\ g^{\mathrm{eff}}_{aN}$}

\def\MJ{{\sc Majorana}}
\def\DEM{{\sc Demonstrator}}
\def\MJD{{\sc Majorana Demonstrator}}

\begin{document}
\title{A Low Energy Rare Event Search with the \MJD}

\newcommand{\ITEP}{National Research Center ``Kurchatov Institute'' Institute for Theoretical and Experimental Physics, Moscow, 117218 Russia}
\newcommand{\lbnl}{Nuclear Science Division, Lawrence Berkeley National Laboratory, Berkeley, CA 94720, USA}
\newcommand{\lanl}{Los Alamos National Laboratory, Los Alamos, NM 87545, USA}
\newcommand{\queens}{Department of Physics, Engineering Physics and Astronomy, Queen's University, Kingston, ON K7L 3N6, Canada}
\newcommand{\uw}{Center for Experimental Nuclear Physics and Astrophysics, and Department of Physics, University of Washington, Seattle, WA 98195, USA}
\newcommand{\unc}{Department of Physics and Astronomy, University of North Carolina, Chapel Hill, NC 27514, USA}
\newcommand{\duke}{Department of Physics, Duke University, Durham, NC 27708, USA}
\newcommand{\ncsu}{Department of Physics, North Carolina State University, Raleigh, NC 27695, USA}	
\newcommand{\ornl}{Oak Ridge National Laboratory, Oak Ridge, TN 37830, USA}
\newcommand{\ou}{Research Center for Nuclear Physics, Osaka University, Ibaraki, Osaka 567-0047, Japan}
\newcommand{\pnnl}{Pacific Northwest National Laboratory, Richland, WA 99354, USA}
\newcommand{\ttu}{Tennessee Tech University, Cookeville, TN 38505, USA}
\newcommand{\sdsmt}{South Dakota School of Mines and Technology, Rapid City, SD 57701, USA}
\newcommand{\usc}{Department of Physics and Astronomy, University of South Carolina, Columbia, SC 29208, USA}
\newcommand{\usd}{Department of Physics, University of South Dakota, Vermillion, SD 57069, USA}  
\newcommand{\ut}{Department of Physics and Astronomy, University of Tennessee, Knoxville, TN 37916, USA}
\newcommand{\tunl}{Triangle Universities Nuclear Laboratory, Durham, NC 27708, USA}
\newcommand{\mpi}{Max-Planck-Institut f\"{u}r Physik, M\"{u}nchen, 80805 Germany}
\newcommand{\tum}{Physik Department and Excellence Cluster Universe, Technische Universit\"{a}t, M\"{u}nchen, 85748 Germany}
\newcommand{\MIT}{Department of Physics, Massachusetts Institute of Technology, Cambridge, MA 02139, USA} 
\newcommand{\williams}{Department of Physics, Williams College, Williamstown, MA 01267, USA}

\affiliation{\uw}
\affiliation{\pnnl}
\affiliation{\usc}
\affiliation{\ornl}
\affiliation{\ITEP}
\affiliation{\usd}
\affiliation{\queens} 
\affiliation{\sdsmt}
\affiliation{\duke}
\affiliation{\tunl}
\affiliation{\uw}
\affiliation{\unc}
\affiliation{\lbnl}
\affiliation{\lanl}
\affiliation{\ut}
\affiliation{\ou}
\affiliation{\ncsu}
\affiliation{\MIT}
\affiliation{\ttu}
\affiliation{\mpi}
\affiliation{\tum}

\author{C.~Wiseman}\email{Corresponding author: wisecg@uw.edu}\affiliation{\uw}
\author{I.J.~Arnquist}\affiliation{\pnnl} 
\author{F.T.~Avignone~III}\affiliation{\usc}\affiliation{\ornl}
\author{A.S.~Barabash}\affiliation{\ITEP}
\author{C.J.~Barton}\affiliation{\usd}	
\author{F.E.~Bertrand}\affiliation{\ornl}
\author{B.~Bos}\affiliation{\sdsmt}\affiliation{\unc}\affiliation{\tunl} 
\author{M.~Busch}\affiliation{\duke}\affiliation{\tunl}	
\author{M.~Buuck}\altaffiliation{Present address: Kavli Institute for Particle Astrophysics and Cosmology, SLAC National Accelerator Laboratory, Menlo Park, CA 94025}\affiliation{\uw}  
\author{T.S.~Caldwell}\affiliation{\unc}\affiliation{\tunl}	
\author{Y-D.~Chan}\affiliation{\lbnl}
\author{C.D.~Christofferson}\affiliation{\sdsmt} 
\author{P.-H.~Chu}\affiliation{\lanl} 
\author{M.L.~Clark}\affiliation{\unc}\affiliation{\tunl} 
\author{C.~Cuesta}\altaffiliation{Present address: Centro de Investigaciones Energ\'{e}ticas, Medioambientales y Tecnol\'{o}gicas, CIEMAT 28040, Madrid, Spain}\affiliation{\uw}	
\author{J.A.~Detwiler}\affiliation{\uw}	
\author{A.~Drobizhev}\affiliation{\lbnl} 
\author{D.W.~Edwins}\affiliation{\usc} 
\author{Yu.~Efremenko}\affiliation{\ut}\affiliation{\ornl}
\author{H.~Ejiri}\affiliation{\ou}
\author{S.R.~Elliott}\affiliation{\lanl}
\author{T.~Gilliss}\affiliation{\unc}\affiliation{\tunl}  
\author{G.K.~Giovanetti}\affiliation{\williams}  
\author{M.P.~Green}\affiliation{\ncsu}\affiliation{\tunl}\affiliation{\ornl}   
\author{J.~Gruszko}\affiliation{\MIT} 
\author{I.S.~Guinn}\affiliation{\unc}\affiliation{\tunl} 
\author{V.E.~Guiseppe}\affiliation{\ornl}	
\author{C.R.~Haufe}\affiliation{\unc}\affiliation{\tunl}	
\author{R.J.~Hegedus}\affiliation{\unc}\affiliation{\tunl} 
\author{R.~Henning}\affiliation{\unc}\affiliation{\tunl}
\author{D.~Hervas~Aguilar}\affiliation{\unc}\affiliation{\tunl} 
\author{E.W.~Hoppe}\affiliation{\pnnl}
\author{A.~Hostiuc}\affiliation{\uw} 
\author{M.F.~Kidd}\affiliation{\ttu}	
\author{I.~Kim}\affiliation{\lanl} 
\author{R.T.~Kouzes}\affiliation{\pnnl}
\author{A.M.~Lopez}\affiliation{\ut}	
\author{J.M. L\'opez-Casta\~no}\affiliation{\usd} 
\author{E.L.~Martin}\affiliation{\unc}\affiliation{\tunl}	
\author{R.D.~Martin}\affiliation{\queens}	
\author{R.~Massarczyk}\affiliation{\lanl}		
\author{S.J.~Meijer}\affiliation{\lanl}	
\author{S.~Mertens}\affiliation{\mpi}\affiliation{\tum}		
\author{J.~Myslik}\affiliation{\lbnl}		
\author{T.K.~Oli}\affiliation{\usd}  
\author{G.~Othman}\affiliation{\unc}\affiliation{\tunl} 
\author{W.~Pettus}\affiliation{\uw}	
\author{A.W.P.~Poon}\affiliation{\lbnl}
\author{D.C.~Radford}\affiliation{\ornl}
\author{J.~Rager}\affiliation{\unc}\affiliation{\tunl}	
\author{A.L.~Reine}\affiliation{\unc}\affiliation{\tunl}	
\author{K.~Rielage}\affiliation{\lanl}
\author{N.W.~Ruof}\affiliation{\uw}	
\author{B.~Sayki}\affiliation{\lanl} 
\author{M.J.~Stortini}\affiliation{\lanl} 
\author{D.~Tedeschi}\affiliation{\usc}		
\author{R.L.~Varner}\affiliation{\ornl}  
\author{J.F.~Wilkerson}\affiliation{\unc}\affiliation{\tunl}\affiliation{\ornl}    

\author{W.~Xu}\affiliation{\usd} 
\author{C.-H.~Yu}\affiliation{\ornl}
\author{B.X.~Zhu}\altaffiliation{Present address: Jet Propulsion Laboratory, California Institute of Technology, Pasadena, CA 91109, USA}\affiliation{\lanl}

\collaboration{{\sc{Majorana}} Collaboration}
\noaffiliation

\begin{abstract}
  The \textsc{Majorana Demonstrator} is sensitive to rare events near its energy threshold, including bosonic dark matter, solar axions, and lightly ionizing particles.
  In this analysis, a novel training set of low energy small-angle Compton scatter events is used to determine the efficiency of pulse shape analysis cuts, and we present updated bosonic dark matter and solar axion results from an 11.17 kg-y dataset using a 5 keV analysis threshold.
\end{abstract}

\maketitle

\section{Introduction}

  The \MJD\ is a neutrinoless double beta decay (\nonubb) experiment, operating an array of P-type point contact high purity germanium (PPC HPGe) detectors, 29.7 kg of which have been enriched to 88\% $^{76}$Ge~\cite{mjd2014}.
  During detector fabrication, surface exposure times were carefully monitored and restricted, resulting in very low levels of cosmogenic backgrounds.
  The world-leading energy resolution, low thresholds, and low electronics noise of the array enable additional searches for beyond-Standard Model physics, including bosonic dark matter~\cite{pospelov} and solar axions.
  
  The initial commissioning data set of 478 kg-d \enrge\ exposure was used to search for bosonic dark matter, solar axions, and other rare events~\cite{mjd2017}.
  Since then, the \DEM\ has amassed a larger exposure ($>$30 kg-y) with roughly two-thirds of it blinded, and several significant hardware and software upgrades to the system have been made, most notably the installation of a second detector module and the completion of the passive shield~\cite{mjd2019}.
  Hence, the work presented here uses improved analysis techniques and nearly a factor of 10 more enriched exposure (11.17 kg-y) with backgrounds reduced by a factor of 4 (to $\sim$0.01 cts/kg-d/keV between 20--40 keV).
  
  Standard data taking for the \DEM\ includes long periods of background (physics) data taking divided into one-hour runs, interspersed with routine $^{228}$Th calibrations used to update energy estimation and pulse shape discrimination parameters.
  The primary \nonubb\ analysis performs the initial processing of the raw data, including  event building, digitizer nonlinearity corrections, energy estimation, and muon and pulser event rejection.
  A low-energy extension of the analysis toolkit has been developed to apply additional techniques to the low S/N region of the data, including threshold measurement, high-frequency noise rejection, wavelet denoising, and a novel slow pulse determination method.
  In these proceedings we describe the slow pulse analysis and its acceptance of fast events as a function of energy, and present updates to the bosonic dark matter and 14.4 keV solar axion searches.

\section{Slow Pulse Discrimination}

  In a PPC, the lithium diffusion process to form the n$^+$ contact creates a $\sim$ 1 mm \textit{dead layer} where the large number of Li atoms effectively cancel out the electric field present in the bulk.
  Ionization (``slow pulse'') events can occur in a region between the dead layer and the bulk material, known as the \textit{transition layer}~\cite{cogent}.
  The \DEM\ reliably triggers on sub-keV events, but it can be difficult to identify distinct features in the rising edge of these very low signal-to-noise waveforms.
  In the 2017 analysis, slow pulses were removed by applying a triangular filter.
  This method works well to 5 keV in the \DEM, but can be improved upon to lower the analysis threshold further.
  Here, we calculate the slowness of each waveform in two steps.
  Waveforms are first wavelet-denoised, and then fit to an exponentially modified Gaussian:
  \begin{align} 
  \begin{split}
    \mathrm{G}(t) = &\frac{A}{2\tau}\ \mathrm{exp} \bigg(\frac{t-\mu}{\tau} + \frac{\sigma^2}{2\tau^2}\bigg) \\
    &\times \mathrm{erfc}\bigg(\frac{\sigma}{\tau \sqrt{2}} - \frac{t-\mu}{\sigma \sqrt{2}}\bigg) + B
  \end{split}
  \end{align}
  This function is a heuristic, but it closely resembles typical PPC waveforms.
  Setting $\tau=-72\ \mu$s gives the model an exponentially decaying tail, matching the typical decay constant of the signal preamplifiers.
  The mean $\mu$ corresponds to the location of the rising edge.
  The amplitude $A$ is proportional to the energy, and the constant offset $B$ accounts for the detector baseline.
  Most importantly, the parameter $\sigma$ is correlated with the slope of the rising edge, making an effective \textit{slowness} parameter.
  This waveform fit can be a computationally intensive task, but shows improved slow pulse rejection efficiency for very low S/N (to 1 keV).


  The method of previous low energy PPC HPGe experiments including CoGeNT~\cite{cogent} and  MALBEK~\cite{malbek} is to use the \textit{fast signal acceptance} as the efficiency correction to the final energy spectrum.
  This is defined by the ratio of fast events in a training set that pass a particular slow pulse cut to a known total, as a function of energy.
  Here, we employ a training set of events taken from routine $^{228}$Th calibrations, updating cut values in time throughout the background dataset.

  
  The strings of HPGe detectors in the \DEM\ allow many possible paths for gamma ray scattering events, and the world-leading energy resolution allows populations of events in calibration data with carefully controlled energies to be studied.
  When the calibration line sources are deployed around each module, we observe the 238.632 keV line from $^{212}$Pb in the $^{228}$Th decay chain.
  Since it is not emitted in cascade with other gammas, a multiplicity-2, sum energy 238 keV event in the array is most likely to be a single Compton scatter in one detector followed by absorption in the second.
  If either detector hit is significantly energy degraded, the sum energy does not contribute to the sum-238 peak, and we can identify a training set of predominantly fast events by selecting events within 3$\sigma$ of the sum energy peak, which we find contains a factor 50 more events above the background.
  
\begin{figure*}
    \centering
    \includegraphics[height=5.6cm]{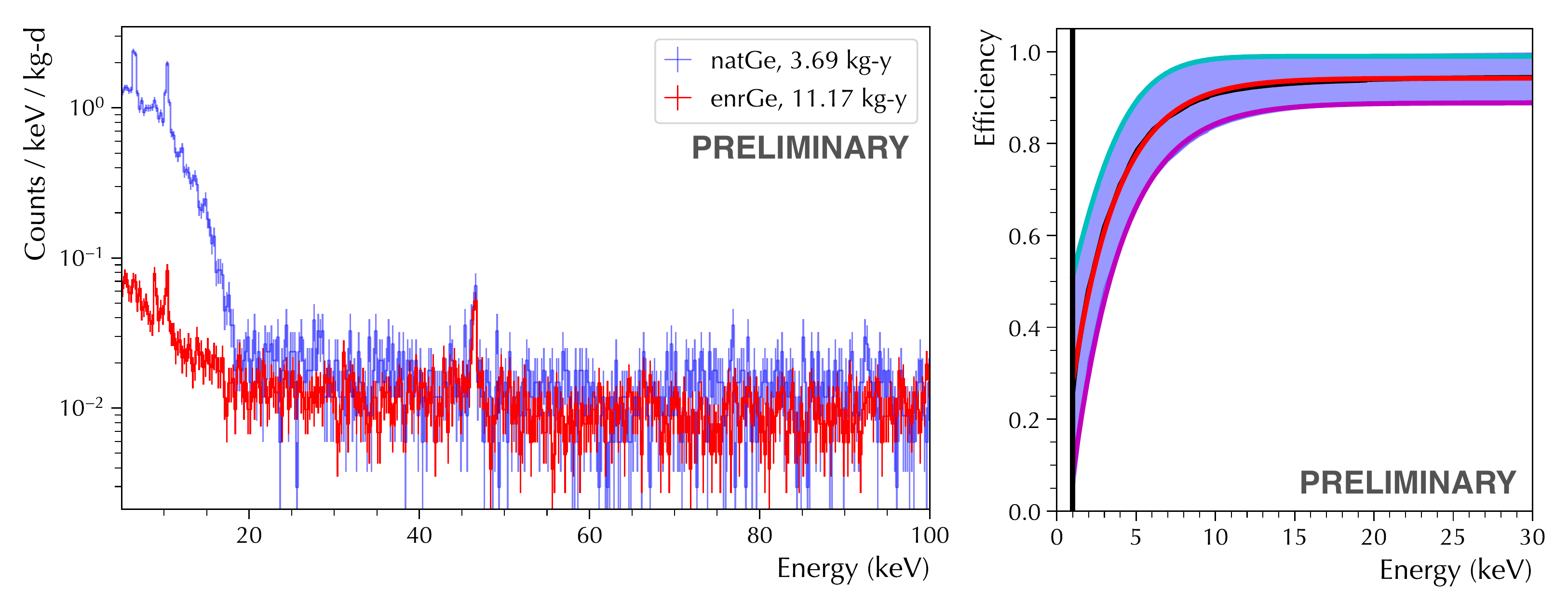}
    \caption{\label{spectrum} (color online)
    \textit{Left:} Enriched and natural spectrum from the 11.17 kg-y dataset. With the increased exposure we also observe a $^{210}$Pb line at 46 keV.
    \textit{Right:} Final efficiency for the \enrge\ detectors.
    The centroid (black) and uncertainty (blue, shaded) are fit to Weibull functions for the centroid (red), upper (cyan) and lower (magenta) limits.}
  \end{figure*}

  The rate of these events is only $\sim$ 1 Hz between all detectors during calibration, and requires we combine all calibration data to obtain a sufficient number to calculate individual detector efficiencies.
  Only one detector in the current analysis is removed from the analysis for having an insufficient number of sum-238 events with hits below 10 keV.
  The low-energy sum-238 detector hit distributions are combined based on the typical rise time of events between 10--200 keV, using $\sigma$ from the waveform fit results, updating the values at every calibration in a database.
  The efficiency is calculated by accepting 95\% of the events between 10 and 100 keV, and then evaluating the acceptance of the cut on the multiplicity-2, sum-238 population below 10 keV.
  The accompanying uncertainty is calculated by a Toy Monte Carlo method where the fit is re-run after varying the number of passing/failing events in each bin according to its Poisson-distributed number of counts.
  Both the centroid and the upper/lower uncertainties are fit to a Weibull distribution, which is used in the rare event search described in the next section.
  Increasing the size of the calibration dataset in the future will reduce this uncertainty.
  The final efficiency is an exposure-weighted sum from each detector.
  The spectrum after the slow pulse cut and final run and channel selection is shown in Figure~\ref{spectrum}, along with the total \enrge\ efficiency curve.
  
  \begin{figure*}
    \centering
    \includegraphics[height=5.1cm]{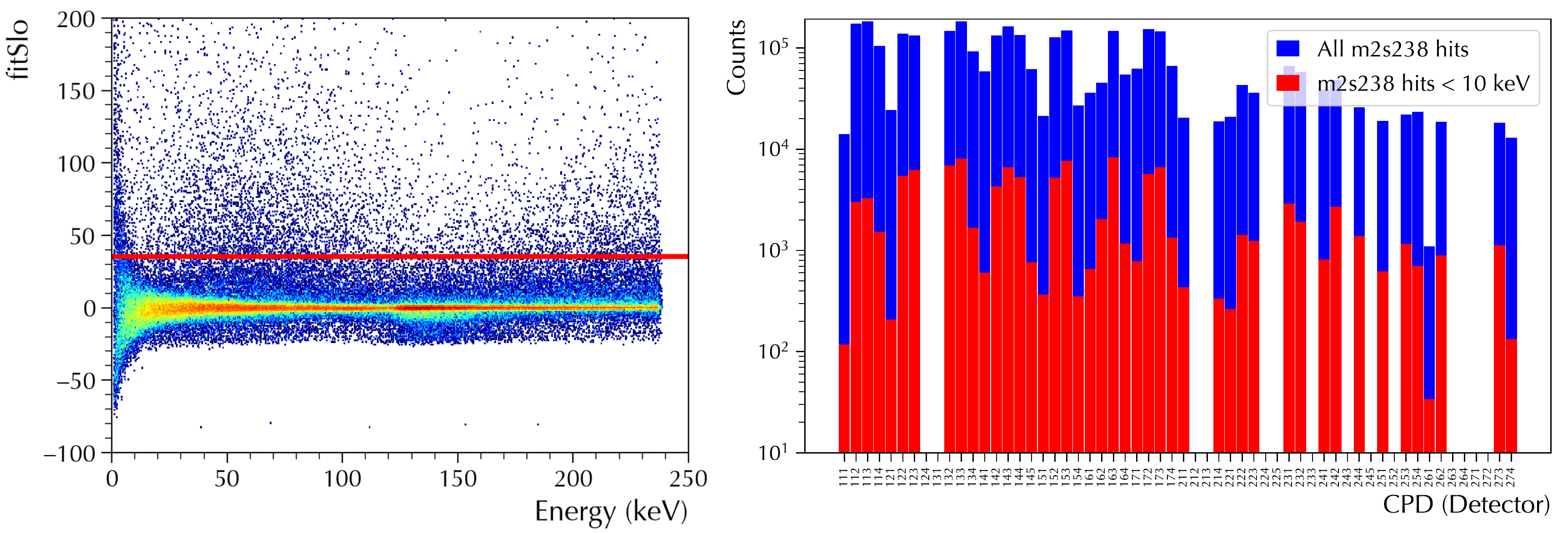}
    \caption{\label{m2s238} (color online)
    \textit{Left:} All low-energy detector hits from the multiplicity-2, sum-238 training population for an example detector (C1P6D3). The cut value (red line) keeps 95\% of events between 10--200 keV.
    \textit{Right:} Rate of multiplicity-2, sum-238 events for all active \enrge\ detectors in the array, which varies considerably with position relative to the calibration sources.}
  \end{figure*}
  
\section{Rare Event Search}

  The current low energy background model for the \DEM\ consists of the tritium beta decay spectrum, a linear continuum, and several spectral peaks modeled as Gaussian.
  The tritium $\beta$-decay has an 18.6 keV endpoint and a 12.32 year half-life.
  The isotope is cosmogenically produced within HPGe crystals and can become the dominant background at low energy if the detector is kept aboveground for a long period.
  Even brief exposures to the cosmic ray flux at high altitudes by transport via airplane can create excessively high tritium levels.
  The manufacture, processing, and storage of the \enrge\ detectors used in the \DEM\ was done to minimize surface exposure time at all points during the manufacturing process~\cite{mjdprocessing}.
  We model the signals of interest from bosonic dark matter and (separately) a 14.4 keV solar axion as a Gaussian peak with a width derived by the \nonubb\ analysis~\cite{mjd2019} and taking a constant 30\% uncertainty over the energy range of interest.
  We perform an unbinned extended maximum likelihood fit of our background model against the data using the RooFit framework~\cite{roofit}.
  Upper limits to 90\% CL on any observable are calculated by a profile likelihood method.
  The fitting method was first used for the 2017 analysis, and the implementation here is nearly identical.
  
\begin{figure*}
    \centering
    \includegraphics[width=0.49\linewidth]{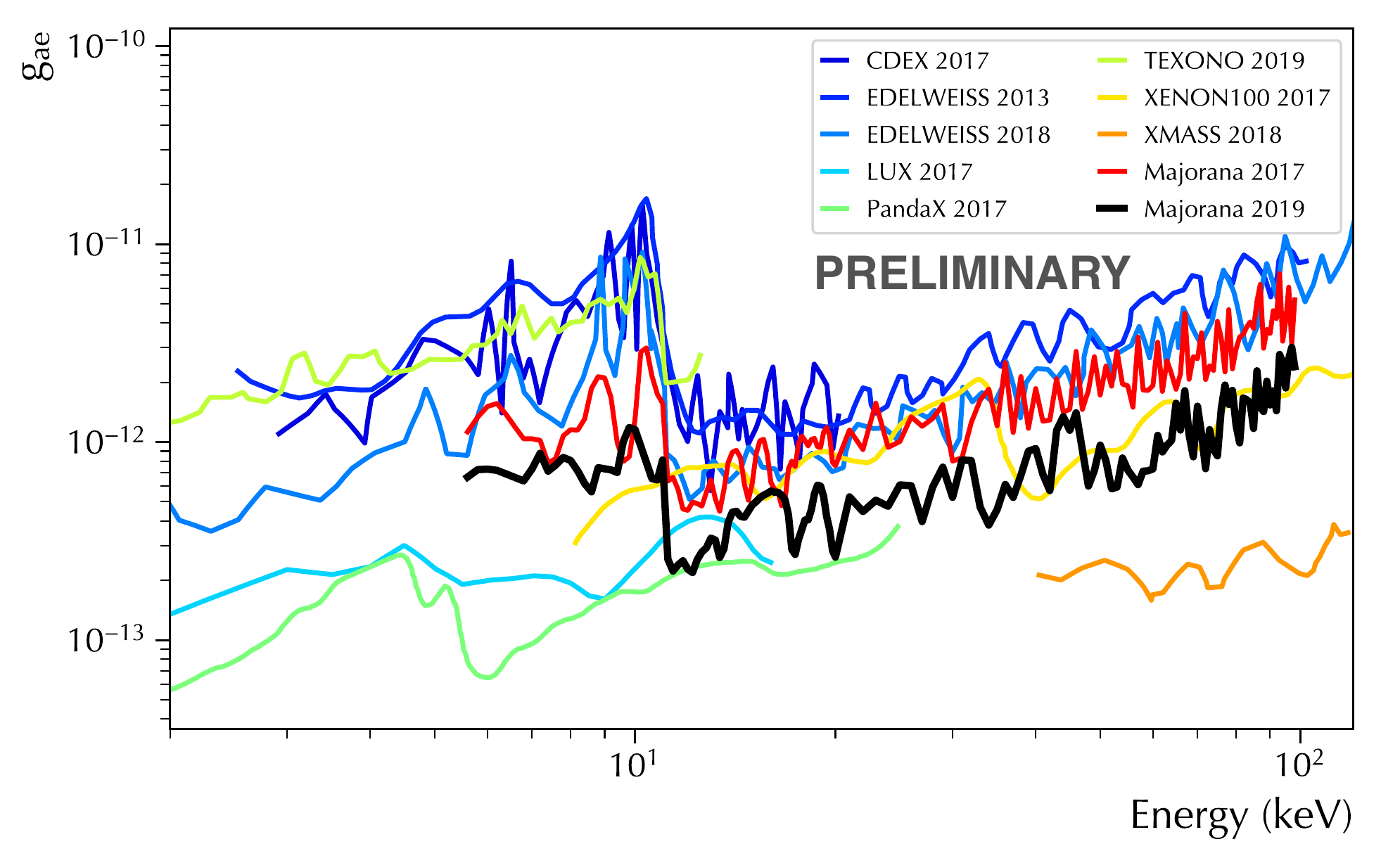}
    \includegraphics[width=0.49\linewidth]{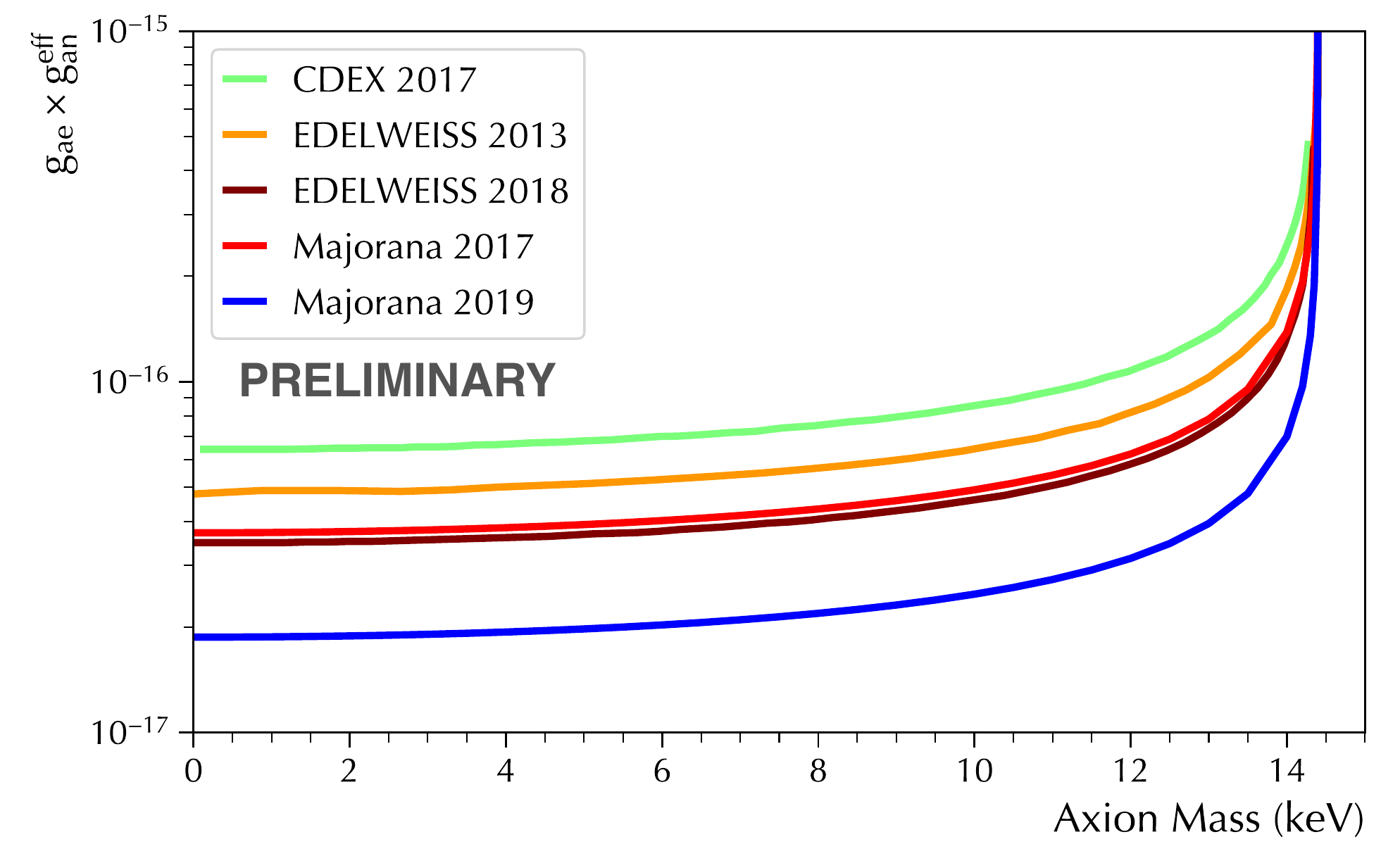}
    \caption{\label{moneyplots} (color online)
    \textit{Left:} Exclusion limits for the Data Set 1--6A analysis (black), plotted against the previous \MJ\ result (blue), and other current experimental limits (see legend).
    \textit{Right:} Exclusion limits for a 14.4 keV solar axion search (blue) shown against previous EDELWEISS and CDEX results.}
  \end{figure*}

  To set exclusion limits for rare event signals, we compare an \textit{expected} number of counts $N_\mathrm{exp}$ to an \textit{observed} number of counts $N_\mathrm{UL}$ obtained from the spectral fit, $N_\mathrm{UL} = A N_\mathrm{exp}$.  
  Here $A$ is the arbitrary coupling constant of interest, \gae\ for bosonic dark matter and \gaegan\ for the solar axion.
  In this analysis, the expected number of counts is obtained from the exposure $MT$ (kg-d), expected flux $\Phi$ (cm$^{-2}$ d$^{-1}$), and cross section $\sigma$ (cm$^2$/kg), each of which is generally dependent on both mass $m$ and total energy $E$.
  The expected fluxes are given in Ref.~\cite{mjd2017}.
  The preliminary results of the upper bounds on bosonic dark matter and solar axion coupling constants are shown in Figure~\ref{moneyplots}.
  
  This material is based upon work supported by the U.S. Department of Energy, Office of Science, Office of Nuclear Physics, the Particle Astrophysics and Nuclear Physics Programs of the National Science Foundation, the Russian Foundation for Basic Research, the Natural Sciences and Engineering Research Council of Canada, the Canada Foundation for Innovation John R.~Evans Leaders Fund, the National Energy Research Scientific Computing Center, and the Oak Ridge Leadership Computing Facility, and the Sanford Underground Research Facility.

\end{document}